\documentclass[global,twocolumn]{svjour_condmat}
\usepackage{graphics}
\journalname{Applied Physics B 82, 207 (2006)}
\begin{document}
\title{Realization of a single Josephson junction for Bose-Einstein condensates }
\author{R. Gati,  M. Albiez, J. F\"olling, B. Hemmerling, M.K. Oberthaler 
}                     
\institute{Kirchhoff Institut f\"ur Physik, University of
Heidelberg, Im Neuenheimer Feld 227, 69120 Heidelberg \\
Fax:+49 6221 549869. E-mail: noisethermometry@matterwave.de}
\date{ }
%
\maketitle
\begin{abstract}
We report on the realization of a double-well potential for
Rubidium-87 Bose-Einstein condensates. The experimental setup
allows the investigation of two different dynamical phenomena
known for this system - Josephson oscillations and self-trapping.
We give a detailed discussion of the experimental setup and the
methods used for calibrating the relevant parameters. We compare
our experimental findings with the predictions of an extended
two-mode model and find quantitative agreement. \\ \\
\noindent \textbf{PACS}: 74.50+r, 03.75.Lm, 05.45.-a
\end{abstract}
\section{Introduction} \label{intro}
One of the most prominent features of quantum mechanics is the
tunnelling of massive particles through classically forbidden
regions. Although tunnelling is a purely quantum mechanical
effect, it can be observed on a macroscopic scale if the system
can be described by two weakly linked (i.e. having a small spatial
overlap) macroscopic wave functions with global phase coherence. A
fundamental physical phenomenon based on macroscopic tunnelling is
the Josephson effect predicted by Brian D. Josephson in 1962
\cite{Josephson}. The first observation of this effect has been
realized with two superconductors separated by a thin insulating
barrier and was reported by Anderson {\em et al.} \cite{Anderson}
just one year after its theoretical prediction. Since then many
experiments with the electronic Josephson junction system have
been performed. The Josephson effect has also found its way to
technological applications such as superconducting quantum
interference device (SQUIDS) which allow to measure weak magnetic
fields with very high precision.

The Josephson effect for neutral superfluids has already been
observed with liquid $^3$He \cite{He3} and $^4$He \cite{He4}
exhibiting the typical current-phase relation. It is characterized
by an alternating current flowing through the central tunnelling
barrier if a constant energy difference between both sides is
applied. In the context of dilute quantum gases such as
Bose-Einstein condensates (BECs) Josephson junction arrays have
been demonstrated \cite{Cataliotti} and only recently a single
weak link was realized and the predicted tunnelling dynamics has
been observed \cite{Albiez}. The main subject of this paper is the
discussion of this experimental setup. We will give a detailed
description of the experimental implementation and the necessary
thorough calibration of system's parameter. We are also going to
discuss the comparison of the obtained data with an extended
two-mode model recently developed by D. Ananikian and T. Bergeman
\cite{Bergeman}.

\section{Basic setup}
\begin{figure}[h]
\center
\resizebox{0.40\textwidth}{!}{%
  \includegraphics{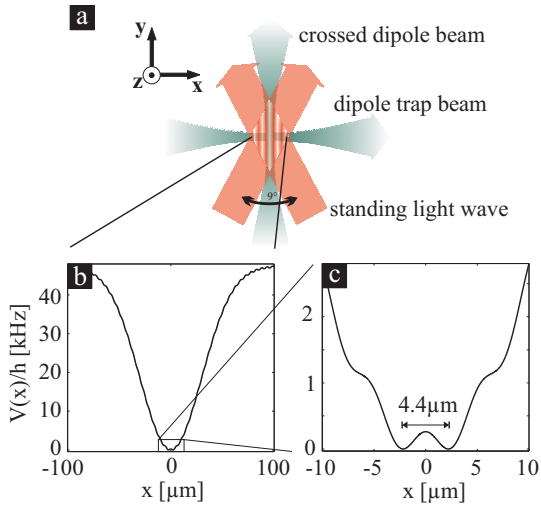}
}
\caption{Realization of a single Josephson junction using optical
dipole potentials. a) The basic setup is given by a crossed beam
dipole trap combined with a periodic light shift potential
implemented with two beams intersecting under an angle of
9$^\circ$. b) On the scale of the three dimensional harmonic
confinement the periodic potential is hardly visible. c) In the
center of the combined potential a clean double-well potential is
realized.}
\label{fig:1}       
\end{figure}

There are many different methods to produce a double-well
potential for Bose-Einstein condensates. The first realized
double-well potential was obtained already in the early days of
BEC using a harmonic confinement created by a magnetic trap and a
focussed blue detuned laser beam generating the tunnelling
barrier. This setup with a well separation of typically 50$\mu$m
was used for the first interference measurements with BECs
\cite{10}. Many other attempts have been made to realize
double-well potentials \cite{double well,double well2,double
well3} but so far only the splitting of condensates into two
independent parts was possible. Recently we succeeded to observe
Josephson dynamics in a double-well potential realized with
optical dipole potentials \cite{Albiez}. The basic idea is the
combination of a three dimensional harmonic confinement and a one
dimensional periodic potential with a large lattice spacing of 5
$\mu$m (see fig.~\ref{fig:1}).

The $^{87}$Rb BEC in our experiment is prepared in a crossed beam
dipole trap consisting of two orthogonal far red detuned Nd-Yag
laser beams (see fig.\ref{fig:1}). A frequency difference of 10MHz
between the two beams is introduced in order to avoid uncontrolled
interferences between both beams. The dipole trap beam is radially
symmetric and has a waist of 60(5)$\mu$m. It provides confinement
in the direction of gravity (in the following discussion
z-direction) and in y-direction. The crossed dipole beam is
radially asymmetric and has a radius at the intersection of
140(5)$\mu$m in z-direction and 70(5)$\mu$m in x-direction. This
asymmetry allows the adjustment of the harmonic confinement along
the x-direction, i.e. the direction of the double-well potential,
without significantly changing the trap frequencies in the other
directions. With a maximum power of approximately 500mW in the
dipole trap beam and 800mW in the crossed dipole beam, we can
achieve a maximum three-dimensional confinement of
$\omega_{x,max}\approx2\pi\times120(6)$Hz, $\omega_{y,max}\approx
2\pi\times170(8)$Hz and $\omega_{z,max}\approx 2\pi\times180(8)$Hz
and a maximum trap depth of approximately 5$\mu$K. The typical
spontaneous emission rate is estimated to be 0.01Hz and thus is
negligible for all our experiments.

The periodic potential is realized by a pair of laser beams with
parallel linear polarization crossing at a relative angle of
$\alpha$=9(1)$^\circ$ as depicted in fig.~\ref{fig:1}(a). The
beams are aligned symmetrically with respect to the crossed dipole
beam, such that the resulting potential is modulated only in
x-direction. The laser wavelength is chosen to be 811nm. The
resulting potential is shown in fig. \ref{fig:1}(b,c) and is given
by
\begin{equation}
V_{dw}=\frac{1}{2} m w_x^2(x-\Delta x)^2+V_0 \cos^2\left(\frac{\pi
x}{d_l} \right)
\end{equation}
where $\Delta x$ is the relative position of the two potentials. For
$\Delta x=0$ a symmetric double-well potential is obtained.

The intensities of all laser beams and the relative distance
between the center of the crossed dipole beam and a maximum of the
periodic potential are very critical parameters thus an active
stabilization is inevitable. The light intensities are stabilized
to better than $10^{-4}$ which allows the generation of small
condensates with small shot-to-shot fluctuations of the final atom
numbers, i.e. 1100(300) atoms. The control of the position of the
crossed dipole beam with respect to the standing light wave is
crucial for the investigation of the dynamics of bosonic Josephson
junctions. The relative shift determines the resulting shape of
the effective double-well and its symmetry. In order to stabilize
the position of the periodic potential, the relative phase of the
two lattice laser beams is controlled with two acousto-optical
modulators. The absolute position of the periodic potential maxima
is deduced by coupling out $2\%$ of the beam intensities and
bringing them to interference under an angle of $6(1)^\circ$. This
leads to an interference pattern with a periodicity of about $8\mu
m$. A precision air slit with a width of 5(1) $\mu$m and a height
of 3mm is placed in the overlap region and is oriented parallel to
the interference fringes. The light intensity transmitted through
the slit, which directly depends on the relative phase between the
two beams, is monitored with a photodiode. The signal of the
photodiode is then locked to a given value using a standard
proportional-integral loop amplifier. The achieved position
stability is deduced by analyzing the position of the split-BEC
images and gives an upper bound for the stability. We obtain a
standard deviation of approximately 100nm (limited by the pixel
size of the camera system), which corresponds to a phase stability
of less than $\pi/50$.

\begin{figure}[h]
\center
\resizebox{0.4\textwidth}{!}{%
  \includegraphics{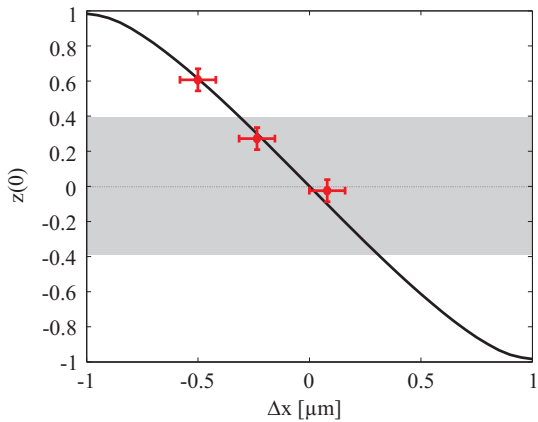}
}
\caption{Preparation of the initial population imbalance. The
filled red circles indicated the measured initial population
imbalance corresponding to the two dynamical situations and the
symmetric case. The condensate in the double-well potential is
prepared by adiabatically ramping up the periodic potential. Since
the symmetry of the effective double-well potential is given by
the relative position $\Delta x$ of the minimum of the crossed
beam dipole trap and a maximum of the periodic potential the
initial population can be adjusted by changing $\Delta x$. The
graph reveals that the initial population imbalance can be created
in the Josephson oscillation regime (indicated by the shading) and
the self-trapping regime. The solid line is the theoretically
predicted imbalance obtained by solving the non-polymomial
Schr\"odinger equation numerically.}
\label{fig:2}       
\end{figure}

In order to initially prepare a given population imbalance, we use
a controlled shift $\Delta x$ of the crossed dipole trap beam
which results in an asymmetric double-well potential, where more
atoms are accumulated in the lower well. This shift is
experimentally implemented with a piezo actuated mirror mount. The
achievable population imbalances and the associated uncertainties
are depicted in fig. \ref{fig:2}. For $\Delta x$=0 the ramping up
of the periodic potential into the condensate prepared in the
harmonic potential leads to a symmetric population of both wells
with vanishing population difference $z=(N_l-N_r)/(N_l+N_r)=0$
($N_{l,r}$ number of atoms in the left and right well,
respectively). A shift of only $\Delta x= 500$nm already
introduces a population imbalance of $z_c=0.39$. As we will show
in the last section this corresponds to the critical imbalance for
our experimental parameters which separates the regime of
Josephson oscillation dynamics, i.e. $z<z_c$ (indicated with gray
shading in fig.\ref{fig:2}) and the self-trapping regime, i.e.
$z>z_c$. Clearly the experimental setup allows the preparation of
both dynamical regimes. After the initial preparation of the BEC
in an asymmetric double-well potential, i.e. $z\neq0$ the dynamics
is initiated by shifting the crossed dipole trap beam back to
$\Delta x=0$. The time scale of 7ms is non-adiabatic with respect
to the typical tunnelling time which is on the order of $50$ms. In
order to understand the experimental observations quantitatively
this finite response time has to be taken into account (see
\cite{Albiez}).

The thorough calibration of all relevant parameters is the
prerequisite for quantitative experiments. The harmonic confinement
and the periodic potential are calibrated independently. The
harmonic trapping frequencies are deduced from standard frequency
measurements of dipolar oscillations. They are excited by reducing
the intensities of the laser beams within less than 500$\mu$s
leading to dipolar oscillations with small amplitudes (few $\mu$m).
These measurements lead to harmonic trapping frequencies of
$\omega_{x}=2\pi\times78(1)$Hz, $\omega_{y}\approx2\pi\times90(1)$Hz
and $\omega_{z}\approx2\pi\times66(1)$Hz.

\begin{figure}[h]
\center
\resizebox{0.45\textwidth}{!}{%
  \includegraphics{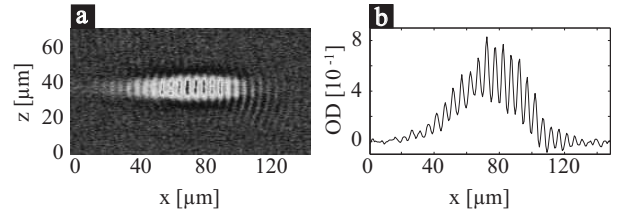}
}
\caption{Calibration of the lattice spacing of the periodic
potential. a) The absorption image of the atomic cloud reveals the
periodic arrangement of pancake BECs. b) The density profile in
x-direction directly shows the lattice spacing $d_l=5.18(9) \mu$m.
From this measurement we can also deduce the optical resolution
(sparrow-criterion) of our imaging system to be 3.2(2)$\mu$m.}
\label{fig:3}       
\end{figure}

Since the tunnelling time is critically dependent on the thickness
and height of the double-well barrier it is crucial to measure
these parameters very accurately. As the width of the barrier is
directly connected to the lattice spacing it can be measured in a
straight forward way. A BEC of $2\times10^4$ atoms is loaded into
a deep optical lattice ($V_0\approx h\times 2$ kHz) superimposed
on a harmonic trap $\omega_{x,y,z}=2\pi\times(5,120,140)$Hz
leading to a periodic array of pancake condensates (see
fig.~\ref{fig:3}). For these parameters the influence of the
shallow harmonic potential in x-direction on the spacing of the
potential is negligible and thus the periodicity can be directly
measured by analyzing the absorption image as shown in fig.
\ref{fig:3}. From that measurement we deduce a lattice constant of
$d_l=5.18(9) \mu$m.

\begin{figure}[b]
\center
\resizebox{0.5\textwidth}{!}{%
  \includegraphics{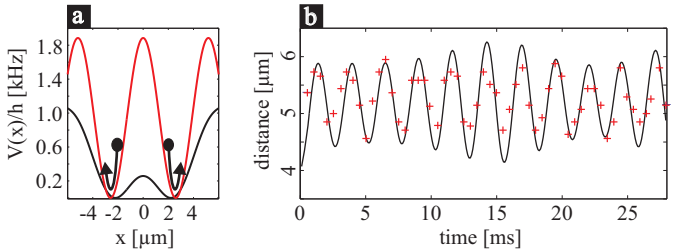}
}
\caption{Calibration of the periodic potential height. a) The BEC
is first loaded into the effective double-well potential (black
solid line). Subsequently the harmonic confinement in x-direction
is switched off and the periodic potential height is increased to
$V_0$. This leads to dipole oscillations in the individual wells
as indicated. b) The relative position of the two BECs in the
optical lattice as a function of time after excitation reveal the
potential height. The experimental data (crosses) are compared to
the numerical simulation (solid line), which has only the lattice
height as a free parameter. We find a potential height of $V_0 =
h\times 412(20)$Hz.}
\label{fig:4}       
\end{figure}

The potential height of the periodic potential is directly
connected to the barrier height. Due to the large well spacings
the standard potential height calibration techniques do not work
and thus we have developed a new method. The potential height is
measured by observing the relative motion of two BECs in the
double-well potential. This motion can be excited by switching off
the harmonic confinement in x-direction and ramping up the barrier
height, i.e. the periodic potential height, by a factor of 5
within 2ms. This procedure leads to a non-adiabatic increase of
the double-well potential spacing $d_{dw}=4.2\mu$m to the lattice
constant of the periodic potential $d_{l}=5.18\mu$m. The time
scale of $2$ms is chosen to reduce the collective excitations in
the transverse direction but leads to very small oscillation
amplitudes of approximately 400nm (see fig.~\ref{fig:4}). However,
this change of the center of mass separation can still be measured
with our optical imaging system which has an optical resolution of
3.2(2) $\mu$m as defined by the Sparrow-criterion. The
experimental data are fitted with the result obtained from the
integration of the non-polynomial Schr\"odinger equation
\cite{Salasnich} and a potential height of $V_0 = h\times
412(20)$Hz is deduced. This corresponds to a barrier height of
$V_b=h\times 263(20)$Hz for the experimentally used harmonic
confinement of $\omega_x=2\pi\times 78$Hz.

\section{Josephson oscillations and self-trapping}

In contrast to all hitherto realized Josephson junctions in
superconductors and superfluids, in our experiment the interaction
between the tunnelling particles plays a crucial role. This
nonlinearity gives rise to new dynamical regimes. Josephson
oscillations, i.e. oscillation of population imbalance and
relative phase of the two condensates, are predicted
\cite{anharmonic_oscill,anharmonic_oscill2,anharmonic_oscill3}, if
the initial population imbalance of the two wells is below a
critical value. The dynamics changes drastically for initial
population differences above the threshold for macroscopic quantum
self-trapping $z_c$ \cite{selftrapping,selftrapping2,Raghavan}
where large amplitude Josephson oscillations are inhibited and the
phase difference increases in time.

The experimental protocol for investigating the dynamics is as
follows: Rubidium atoms are precooled in a standard TOP trap,
transferred into the crossed beam dipole trap and evaporatively
cooled to degeneracy by lowering the light intensity. Finally the
dipole laser beams and the optical lattice beams are ramped to the
desired values. This sequence creates two weakly linked BECs inside
an asymmetric double-well potential with well defined asymmetry and
barrier height. The dynamics is initiated by shifting the crossed
dipole beam realizing a symmetric double-well potential. After a
given evolution time the potential barrier is suddenly ramped up and
the dipole trap beam is switched off. This results in dipole
oscillations of the atomic cloud around two neighboring minima of
the periodic potential as used for the calibration of the periodic
potential height (see fig.\ref{fig:4}). The atomic distribution is
imaged at the time of maximum separation using absorption imaging
techniques. This protocol has been used for the first demonstration
of the transition between Josephson oscillations and macroscopic
self-trapping in a single weak link \cite{Albiez}. In the following
we shall show that the recent work by D. Ananikian and T. Bergeman
\cite{Bergeman} allows to understand the experimental observation
quantitatively with a relatively simple two-mode model.

\begin{figure}[h]
\center
\resizebox{0.35\textwidth}{!}{%
  \includegraphics{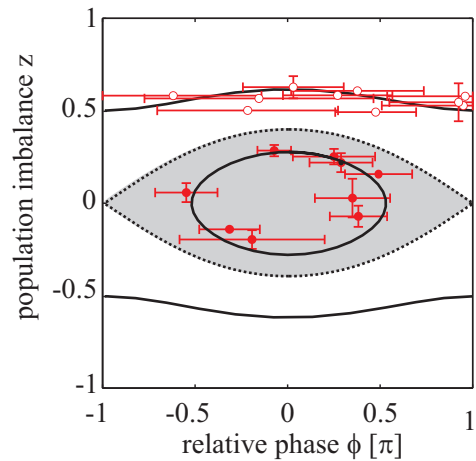}
}
\caption{Comparison of the experimentally obtained phase plane
trajectories to the predictions of the extended two-mode model
(solid lines). The Josephson oscillation regime (gray shaded
region) is characterized by closed trajectories (filled circles).
The separatrix, which is represented by the dashed line,
constitutes the transition to the self-trapped regime (open
circles). It is important to note that there are no free
parameters. Thus the improved two-mode tunnelling model explains
our observation quantitatively.}
\label{fig:5}       
\end{figure}

The standard two-mode approximation assumes a weak link, i.e. the
wave function overlap is negligible \cite{Raghavan}. For our
experimental parameters this is not strictly true. S. Giovanazzi
{\em et al.} \cite{Giovanazzi} have already included the leading
term of the correction to the simple two-mode model but recently D.
Ananikian and T. Bergeman managed to include all terms and present
their improved two-mode model in \cite{Bergeman}. They have derived
differential equations for the basic two-mode parameters the
population difference $z$ and the relative phase $\phi$ between the
two condensates. We will not elaborate on the theoretical
description any further and refer the reader to the reference
\cite{Bergeman} for details. Here, we report on the very good
agreement of the prediction of this model with our experimental
observation concerning the critical population imbalance
distinguishing between the two different dynamical regimes. The
simple constant tunnelling model predicts for our experimental
parameters a critical population difference $z_c=0.23$, but for this
imbalance experimentally still Josephson oscillations are observed.
From the experimental data we deduce $z_{exp}= 0.38 (8)$ which is in
very good agreement with the prediction of the improved two-mode
model giving $z_c=0.35$. This agreement is clearly revealed in
fig.\ref{fig:5} where the phase plane portrait predicted by the time
varying tunnelling model (solid lines) and the experimental data
(circles) are shown. It is important to note that there are no free
parameters in this graph, which is in contrast to the earlier
reported phase-plane portrait \cite{Albiez}, where the critical
population difference was a free fit parameter.

Thus the observed dynamics can be understood in a simple two-mode
model, but a time dependent tunnelling rate has to be taken into
account. This finding is an important prerequisite for further
investigations of more complex systems which are build up by
Josephson junctions. One interesting direction is the controlled
realization of many coupled weak links in two dimensions where the
topology could have a big influence on the dynamics \cite{Andrea}.
Also a very intriguing route is the investigation of the influence
of the residual thermal cloud on the coherent dynamics
\cite{Stringari}. Here, a completely new regime can be reached
with the BEC system since the thermal cloud and thus the
dissipation can be controlled.

We wish to thank Tom Bergeman, Andrea Trombettoni and Augusto Smerzi
for very stimulating discussions. We would also like to thank Matteo
Cristiani and Stefan Hunsmann for their contributions to the
experimental setup. This work was funded by Deutsche
Forschungsgemeinschaft Schwerpunktsprogramm SPP1116 and by
Landesstiftung Baden-W\"urttemberg - Atomoptik. R. Gati thanks the
Landesgraduiertenf\"orderung Baden-W\"urttemberg for the financial
support.

%

\end{document}